# On the Impact of Guest Traffic in Open-Access Domestic Broadband Sharing Schemes

Sotiris Lenas, Vassilis Tsaoussidis
Democritus University of Thrace
{slenas, vtsaousi}@ee.duth.gr

Srikanth Sundaresan
ICSI, Berkeley
srikanth@icsi.berkeley.edu

Arjuna Sathiaseelan, Jon Crowcroft
University of Cambridge
{arjuna.sathiaseelan, Jon.Crowcroft}@cl.cam.ac.uk

*Abstract*—Open-access domestic broadband connection sharing constitutes a voluntary practice that is associated with societal, economic and public-safety benefits. Despite this fact, broadband subscribers are usually hesitant to freely share their broadband connection with guests for a multitude of reasons; one of them being sharing their network might hinder their own broadband quality of experience. In this paper, we investigate experimentally the impact of uplink guest traffic on the sharer's broadband quality of experience under both generic and broadband-sharing-specific packet scheduling policies. Both guest-user traffic and access point profiles employed in our study are developed by analyzing real-world traffic traces and measurements, captured from actual broadband sharing networking environments. Our results validate the suitability of hybrid packet scheduling policies for broadband sharing schemes and show that only a few dozen kilobytes per second of uplink guest traffic can be tolerated by sharers without hampering their broadband quality of experience. In this context, we show that the selection of the most appropriate packet scheduling policy for broadband sharing, as well as its respective configuration, depend largely on the capacity of the broadband connection and the policy's packet-dropping behavior on guest traffic.

I. INTRODUCTION

During the last few years, domestic broadband connection sharing has earned a key role in the broad area of pervasive communications with the spread of Internet sharing services. Recently, new proposals and initiatives have been put forward to extend the original scope of broadband connection sharing as a framework for providing free Internet access [1] [2]. In line with the "The Internet is for Everyone" notion introduced by Vint Cerf [3], free Internet access is associated with multiple benefits, such as reducing the consequences of digital exclusion faced by less-privileged people [4], supporting emergency services, providing ubiquitous networking, and benefiting, both, business and economic growth.

The fundamental idea behind these proposals and initiatives is to consider broadband sharing as a resource pooling service. Through this service, the unused capacity of a broadband connection could be leveraged to carry the traffic generated by *guest users*. Practical examples of such open-access broadband connection sharing frameworks and schemes have been demonstrated in the context of Open Wireless Movement [2] and Public Access WI-FI (PAWS) project [5].

Following the principles of User-Provided Networks (UPNs), in which the consumer of a service becomes the provider, the concept of broadband connection sharing comprises two main entities: i) *home users*, who act as micro-providers by building an access network and sharing their Internet connection and ii) *guest users*, who wish to freely access the Internet by exploiting the available unused capacity.

The network dynamics developed between these two entities are rather complex considering the diverse set of requirements that need to be met. From *home users'* perspective, broadband connection sharing schemes should preserve sufficient quality of experience for *home users*, similar to the one they enjoyed prior to sharing their connection. Unlike *home users*, *guest users* may not expect such high-quality service; rather, in line with the philosophy that governs domestic broadband open-access services, a lower level of service should be satisfactory.

Both the quality of network experience enjoyed by *home users* and the amount of resources allocated to *guest users* are directly affected by the respective scheduling and dropping policies. Currently, Quality of Service (QoS) Differentiated Services Code Point field of the IP header is marked/processed entirely at the Broadband Remote Access Server (BRAS). Since, BRASes are controlled exclusively by Internet Service Providers (ISPs), it becomes apparent that in the context of broadband connection sharing schemes some sophisticated traffic control should also be performed at the edges of the network (i.e. at the access point of the broadband connection).

Therefore, in this paper, our aim is to explore whether the application of either generic or broadband-sharing-specific packet scheduling policies at access points can successfully mitigate the occasionally contradictive set of requirements developed between *home* and *guest users*. We achieve that by investigating experimentally the impact of guest-user uplink traffic on the sharer's broadband quality of experience for each one of the examined policies, under different access point profiles and increasing volumes of guest traffic.

The reason we focus solely on the uplink traffic is twofold: First, broadband connection commercial packages offer abundant downlink bandwidth, several orders of magnitude greater than the uplink bandwidth. In this regard, guest traffic can be easily regulated in the downstream direction with a simple rate-limiting policy. Second, broadband subscribers control over downlink traffic is limited. Access points are the last stations where admission control and QoS policies are applied in the downstream direction, whereas uplink traffic can be regulated entirely at the edges of ISP network.

One key aspect of the present work is that it relies for the experimental investigation on traffic traces and usage data, collected from a real-world broadband sharing trial. Both datasets are analyzed to produce guest-user traffic and access point profiles that are employed to drive the simulations. Overall, our results validate the suitability of hybrid packet scheduling policies for open-access domestic broadband sharing schemes and show that only a few dozen kilobytes per second of uplink guest traffic can be tolerated by sharers without hampering their broadband quality of experience. In this context, we show that the selection of the most appropriate packet scheduling policy for broadband sharing, as well as its respective configuration, depend largely on the capacity of the broadband connection and the policy's packet-dropping behavior on guest traffic.

The remainder of this paper is structured as follows: In Section II, we review the related work. The broadband sharing trial details along with our trace collection and traffic identification methodology are reported in Section III. In Section IV, we present our findings from analyzing the collected usage data. In Section V, we elaborate on the modeling process and present the selected distributions set for modeling guest uplink traffic. In Section VI, we describe the simulation model, elaborate on the experimental settings and present the simulation results. In Section VII, we summarize our conclusions and list future work.

## II. RELATED WORK

Several commercial and non-commercial initiatives have already explored sharing users' broadband connection via various wireless technologies. FON [6], Open Garden [7], OpenSpark [8] and Open-Mesh [9] are among the most popular broadband sharing schemes currently operating across the globe. Although those schemes are gaining worldwide acceptance, their design and deployment philosophy of extending users' paid services confines their operation as free and open-access broadband connection sharing platforms.

Unlike the aforementioned initiatives, Open Wireless Movement [2] and PAWS [5] introduce a new Internet access paradigm, based on a Less-than-Best-Effort (LBE) access model, that enables *guest users* to exploit the spare capacity of domestic broadband networks. In particular, both schemes adopt an approach of community-wide participation, where broadband customers are enabled to donate a portion of their broadband Internet connection for use by fellow citizens.

Regarding the deployment practices for LBE service at the edges of ISPs' network, the proposed approaches can be classified into two generic categories. The first category includes all non-intrusive transport protocol approaches [10], such as the Low Extra Delay Background Transport protocol (LEDBAT) [11]. These approaches are employed on a *per-application* basis and are mainly used to transfer certain application data as background traffic without hampering the network performance of other delay-sensitive Internet applications run by the user. Apart from their typical role though, their use in resource sharing scenarios has also been considered. In particular, the authors of [12] explore the capabilities of LEDBAT to carry non-commercial traffic by exploiting unused 4G satellite link capacity. The second category includes approaches that are based on specialized packet scheduling policies. These are employed on a *per-gateway* basis and their main purpose is to provide *guest users* with a lower-priority access to the network resources in contrast to priority level offered to typical subscribers.

In this context, Psaras et al. introduce UPN Queuing (UPNQ) [13], a packet-scheduling algorithm based on the non-preemptive priority queueing scheme (PQ). UPNQ regulates the service rate of guest traffic by taking into account its percentage impact ($k$) on *home users'* average queuing delay. The authors show, both analytically and experimentally, that by employing UPNQ, a small amount of guest traffic can be served with statistically zero impact on the network performance enjoyed by *home users*.

As initially shown in [14], this percentage-based approach entails a significant drawback; it does not capture any quantitative characteristics of the additionally imposed queueing delay. For example, although, under some certain traffic load, the percentage impact on *home users'* queueing delay might be identical in two different access networks, the respective temporal impact may differ by several milliseconds.

To overcome this drawback, the authors of [15] introduce Hybrid Packet Scheduling Scheme (HPSS). HPSS relies on *home users* queueing delay tolerance level (*HPSS target delay impact*) for its operation. In particular, *HPSS target delay impact* is used as the upper bound of the additional queueing delay imposed to home-user traffic by guest traffic. Based on this *target delay impact* value, HPSS determines analytically the *HPSS capacity threshold*, which corresponds to a particular bandwidth availability point. This threshold reflects the point based on which HPSS selects its *modus operandi* and employs the most appropriate scheduling policy between priority queueing and class-based WFQ. Practically, for access networks with maximum capacity exceeding the *HPSS capacity threshold*, HPSS employs a class-based WFQ policy, in order to guarantee a minimum service rate for *guest users*. The amount of resources allocated to serve *guest users* is estimated analytically. For access networks whose maximum capacity is below that *HPSS capacity threshold*, HPSS employs a non-preemptive PQ policy between home and guest traffic. In this case, unlike UPNQ, HPSS employs a temporal-based packet-scheduling algorithm to confine the additional delay imposed on *home users* average queuing delay below the *HPSS target delay impact* value.

Although methods of both categories can be used supposedly for sharing-resource purposes, in practice, forcing *guest users* to use either non-intrusive transport protocols or specific-purpose applications is more challenging and might as well discourage them from actually using the service. What would be more practical is to transparently control system dynamics between home- and guest-traffic by employing appropriate packet scheduling policies at access points, placed at network edges. That said, we focus on the two queueing methods described in the second category, namely UPNQ and HPSS. We explore their efficiency in sharing domestic broadband network resources to *guest users* and compare them against several other common queueing approaches.

## III. BROADBAND SHARING TRIAL DETAILS

Our broadband sharing trial took place over a seven-month period. Its purpose was to understand the technical constraints of providing free Internet access by sharing existing broadband subscribers' connections. Seventeen broadband sharing gateways were deployed in total, eight of which were used by 15 *guest users* to access the Internet. The nine remaining gateways served exclusively as measurement points.

Each gateway carried out daily measurements of the throughput and latency experienced by each sharer's broadband network. Those measurements are used for evaluating the performance of each access point. The

Fig. 1. Broadband sharing network architecture and data capturing infrastructure.

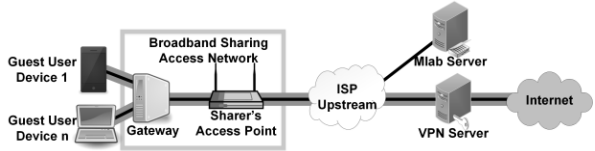

following data were collected:

i) Throughput: Upload and download throughput (Mb/s) was sampled with *netperf* [16] four times a day in order to observe potential bandwidth fluctuations in ISP networks.

ii) End-to-End RTT *(e2ertt):* A set of servers were *pinged* every 10 minutes and the corresponding round-trip-times (RTT) measurements (ms) were collected.

iii) Last-Mile RTT (lmrtt, *ulrttdw and ulrttup*): *lmrtt* represents last mile latency (ms) under such network load, in which network queues were mostly empty, whereas *ulrttdw* and *ulrttup* represent the downstream and upstream last mile latency (ms), under heavy network load. Heavy network load was imposed by running dummy bitrate tests in the background. The Broadband Internet Service Benchmark (*BISmark*) tool [17] was used to obtain these values on a 2-hour interval basis.

Finally, guest-user traffic traces from each individual broadband sharing network were collected on the VPN server (see Fig. 1). During the trials a total of 29GB traffic (10GB upload, 19GB download) were generated by 15 *guest users*.

## IV. DOMESTIC ACCESS POINT PROFILES

An access point profile is mainly characterized by four essential properties: i) Downlink capacity, ii) uplink capacity, iii) downlink queue size and iv) uplink queue size. In order to model these four properties, we analyze data collected from all different broadband sharing networks.

Each one of the 17 access point profiles is abbreviated in Fig. 2 by AP, followed by a corresponding identification number. More specifically, in Fig. 2(a), we show the downlink capacity of each access point, along with its corresponding downlink queue-size values range, whereas in Fig. 2(b), we show the respective values for the upstream direction.

### A. Access Point Throughput Capacity

In order to accommodate the throughput variations throughout different time periods, we use as each access point's throughput capacity the average of the respective downlink and uplink values observed by *netperf*. The throughput capacity results for each access point presented in Fig. 2(a) and Fig. 2(b) for the downstream ($Capacity_{dw}$) and upstream ($Capacity_{up}$) direction, respectively, are used for the rest of the analysis and simulations.

### B. Access Point Uplink and Downlink Queue Size

Access point downlink and uplink queue sizes are determined based on *lmrtt*, *ulrttdw* and *ulrttup* values. In particular, we quantify the buffering effect, in kilobytes (KB),

Fig. 2. Access point profiles produced from the analysis of data collected during the trial. Boxes show interquartile ranges and the median, with the whiskers showing the 10[th] and 90th percentiles.

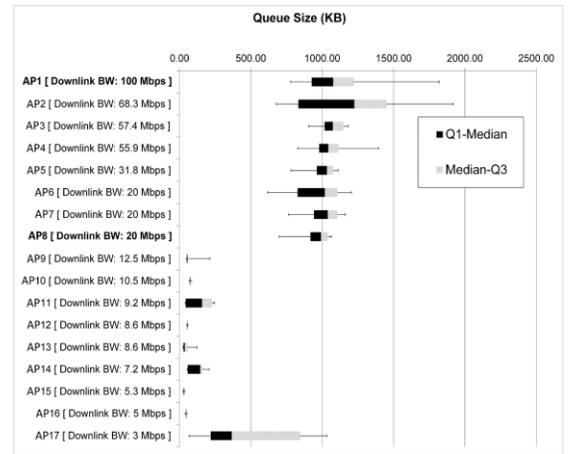

(a) Downlink capacity of each access point along with its corresponding downlink queue size range

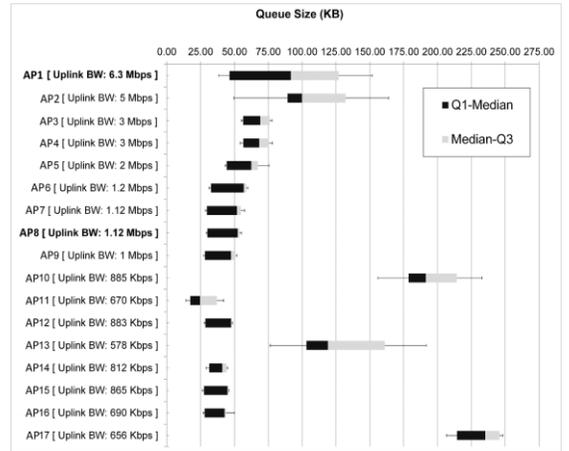

(b) Uplink capacity of each access point along with its corresponding uplink queue size range.

per each traffic direction as follows:

$$buffEffect_{dw} = \frac{(ulrttdw - lmrtt) * Capacity_{dw}}{8} \quad (1)$$

$$buffEffect_{up} = \frac{(ulrttup - lmrtt) * Capacity_{up}}{8} \quad (2)$$

As depicted in Fig.2, the observed downlink median queue sizes for AP1-AP5, which are equipped with fiber-optic links, are close to 1.1MB, whereas the respective uplink median queue sizes are between 60 and 100KB. Regarding the rest of the access points, results for the downstream direction can be divided into two sub-categories. Access points that employ high-capacity conventional Asymmetric Digital Subscriber Line (ADSL) links (i.e. 20Mbps) present downlink median queue sizes close to 1MB, whereas, most of the access points that employ medium-to-low capacity ADSL links (i.e. 3–12.5Mbps) present downlink median queue sizes close to 75KB. The respective uplink median queue sizes for all access points employing conventional links are approximately 50KB.

Some slight deviation from the aforementioned median queue sizes are observed across all cases; this is reasonable given the different broadband connection providers and various access point types (i.e. brands) considered in the study. Furthermore, with respect to AP10, AP11, AP13 and AP17, their median queue sizes deviate significantly from the respective median of other access points with similar capacity profiles. This could be attributed either to a tweaked access point configuration or to some error introduced by *BISmark*.

## V. MODELING GUEST TRAFFIC

### A. The Broad Picture

Out of the 15 *guest users* that took part in the trial, four of them were responsible for the largest part of the traffic, with one of them — the *primary user* — being responsible for 75% of the total guest traffic. Therefore, there was enough evidence to create four distinct guest-user traffic profiles, which in turn allows us to enhance our simulations with sufficient diversity.

In order to comprehend better the scientific properties of the collected traffic traces for these four *guest users* and associate them with qualitative characteristics of well-known statistical distributions, we analyze those traces by following a flow-level analysis approach. In particular, we consider the following flow-level characteristics: i) flow inter-arrival time – the time interval between two consecutive flow arrivals, measured in seconds, ii) flow size – the total number of bytes transferred during a flow and iii) flow duration – the time between the start and the end of a flow, measured in seconds.

For each one of the four *guest users*, we tried to optimally fit the three corresponding empirical distributions, produced by the respective flow-level characteristic samples, to a wide range of typical statistical distributions. Three different statistical tests were employed to evaluate goodness-of-fit in each of these cases, namely the Kolmogorov-Smirnov, Anderson-Darling and Chi-Squared tests. The majority of these tests returned negative results. Various data transformations, such as $\frac{1}{x}$ and $\sqrt{x}$, and combinations of statistical distributions were also employed, achieving poor-to-moderate results. In order to compensate for these discrepancies, we empirically evaluated the goodness-of-fit of various statistical distributions on the sample data.

### B. Empirical Analysis

We assess the goodness-of-fit of various statistical distributions on the sample data using P-P plots. The statistical distribution presenting the smallest overall deviation is chosen as the best-fit. In each of the three panel rows of Fig. 3, we present, respectively, the P-P plots for inter-arrival times, flow sizes and flow durations. Panels (a), (c) and (e) present the P-P plots regarding the *primary user*. In order to have a good grasp of the traffic properties in both ends of the four investigated cases, the panels of the second column present the P-P plots regarding the *fourth guest user*, who generated the least amount of traffic. The respective P-P plots for the other two *guest users*, which were also produced in the course of the study, present similar behavior to those depicted in Fig. 3. For the sake of clarity, each panel depicts the empirical distribution line, along with the plots of the five statistical distributions that present the smallest overall deviation.

*1) Flow Inter-arrival times*

Our sample analysis regarding flow inter-arrival times

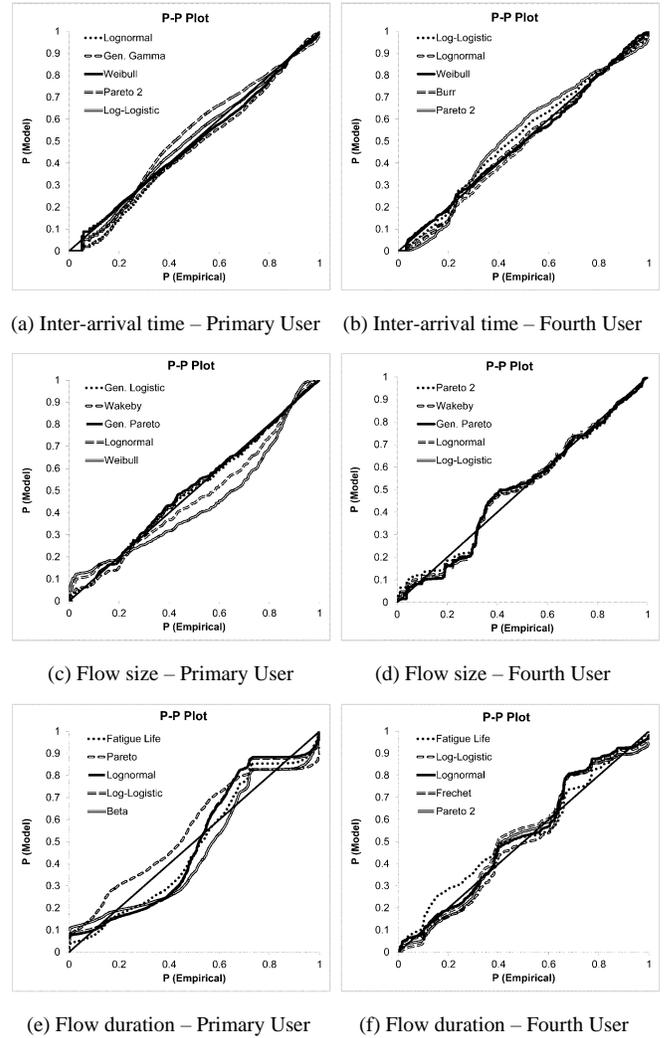

Fig. 3. P-P plots comparing the empirical distribution of inter-arrival time, flow size and flow duration characteristics with various statistical distributions.

(a) Inter-arrival time – Primary User  (b) Inter-arrival time – Fourth User

(c) Flow size – Primary User  (d) Flow size – Fourth User

(e) Flow duration – Primary User  (f) Flow duration – Fourth User

shows that in both cases *Weibull* distribution matches the empirical distribution line well.

More specifically, in Fig. 3(a), although both *Log-logistic* and *Weibull* distributions present similar overall behavior, *Weibull* seems to have a slight advantage over *Log-logistic* distribution at the left tail. In particular, there are many points of the *Log-logistic* distribution in the range of 0 to 0.2, which present considerably larger deviation from the empirical line over the respective *Weibull* distribution points. In Fig. 3(b), *Burr*, *Lognormal* and *Weibull* distributions present similar behavior, with *Lognormal* and *Burr* distributions having larger deviation over the empirical distribution line at the left tail. The rest of the distributions presented in Fig. 3(a) and Fig. 3(b) are, evidently, not appropriate matches.

Furthermore, modelling flow inter-arrival times with Weibull distribution confirms further recent studies for modelling TCP flow inter-arrival times (e.g. [18]) in networking environments similar to ours.

*2) Flow sizes*

It is clear from Fig. 3(c) that *Gen. Logistic* and *Gen. Pareto* distributions present almost identical behavior with minor deviations from the empirical distribution line, rendering them both possible candidates for modeling flow sizes. The results depicted in Fig. 3(d) show that all five distributions present similar light-tailed behavior. The same results also suggest that *Lognormal* and *Gen. Pareto* distributions perform slightly better than the other distributions. *Generalized Pareto* is the common denominator in both cases; therefore, it is selected as the preferred distribution for modeling flow sizes. Pareto-type distributions have also been used in previous studies for modeling TCP flow sizes (e.g. [19]).

*3) Flow durations*

The plots presented in Fig. 3(e) and Fig. 3(f) clearly demonstrate the inability of the presented distributions to sufficiently match the respective flow duration empirical distribution. *Lognormal* distribution seems to perform slightly better in comparison to the rest of the examined distributions. This behavior remains consistent across all four examined cases. Due to this fact, *Lognormal* is selected as the distribution of choice for modeling flow durations.

*C. Empirical Analysis Results*

Table I presents the four guest-user profiles, as produced by the empirical analysis. In particular, it shows the statistical distribution parameters values used for modelling each flow characteristic for all four guest-user profiles. Two main conclusions can be drawn based on the results presented in Table I: First, the vast majority of the captured guest traffic is composed by mice flows. This is clearly indicated by the *Generalized Pareto* parameters values. Second, based on an overall evaluation of the presented values, the average throughput produced by the guest-user profiles is in the range of 1 to 5 KBps, with peak values in the range of 5 to 30 KBps.

*D. Experimental Validation of Empirical Analysis Results*

It is clear from the analysis performed so far that it is not possible to fit perfectly the guest traffic samples for inter-arrival times, flow sizes and flow durations to some standard statistical distributions. In order to validate the suitability of the distributions selected in section V.B, we compare experimentally the uplink traffic properties generated by the empirical distributions (i.e. distributions formed by the raw data) versus the corresponding selected *Weibull*, *Generalized Pareto* and *Lognormal* distributions, for all four guest-user profiles, presented in Table I.

For this purpose, a flow-based traffic generator for TCP flows of specific size and duration over different time intervals was developed in NS2. The following metrics are considered in our validation tests: number of generated packets, number of received packets, average throughput, max throughput (to capture traffic burstiness), and average end-to-end delay.

A simple network topology consisting of two nodes is used for the validation experiments, with the first node being the source and the second one the destination. A Droptail queue is also attached to the source node. Simulation time is set to one hour and 100 simulation runs are completed per modeling case and per guest-user profile.

Table II presents both per metric and per guest-user profile comparison results, which show that in all cases average fitting errors are less than 5%.

VI. EVALUATING THE IMPACT OF GUEST TRAFFIC ON THE

TABLE I. USER PROFILE DISTRIBUTIONS PARAMETERS

| Guest-user Profile | Flow Characteristics | Parameters [a] |
|---|---|---|
| 1 | Inter-arrival time<br>Size<br>Duration | $\alpha = 0.27, \beta = 0.4$<br>$\kappa = 0.59, \sigma = 544, \mu = 353$<br>$\sigma = 2.62, \mu = 1.03$ |
| 2 | Inter-arrival time<br>Size<br>Duration | $\alpha = 0.31, \beta = 0.53$<br>$\kappa = 0.77, \sigma = 1108, \mu = 203$<br>$\sigma = 1.83, \mu = 1$ |
| 3 | Inter-arrival time<br>Size<br>Duration | $\alpha = 0.4, \beta = 0.19$<br>$\kappa = 0.12, \sigma = 1473, \mu = 471$<br>$\sigma = 1.18, \mu = 0.9$ |
| 4 | Inter-arrival time<br>Size<br>Duration | $\alpha = 0.38, \beta = 9.58$<br>$\kappa = 0.59, \sigma = 620, \mu = 42$<br>$\sigma = 2.45, \mu = 1.97$ |

[a] Statistical distribution parameters values used for modeling the respective flow characteristic (i.e. *Weibul* for inter-arrival times, where $\alpha$ and $\beta$ are the respective shape and scale parameters, *Generalized Pareto* for flow sizes, where $\kappa$, $\sigma$ and $\mu$ are the respective shape, scale and location parameters, and *Lognormal* for flow durations, where $\mu$ and $\sigma$ are the respective mean and standard deviation distribution parameters.)

TABLE II. AVERAGE FITTING ERROR BETWEEN EMPIRICAL AND SELECTED DISTRIBUTIONS

| Metrics \ User Profile | 1 | 2 | 3 | 4 |
|---|---|---|---|---|
| Generated Packets | 1.62% | 0.82% | 4.99% | 0.54% |
| Received Packets | 1.62% | 0.82% | 4.99% | 0.54% |
| Avg. Throughput | 1.75% | 0.82% | 5% | 1.15% |
| Max throughput | 3.15% | 0.13% | 4.54% | 3.08% |
| E2E Delay | 0.27% | 0.01% | 0.12% | 0.1% |

## QUALITY OF EXPERIENCE OF HOME USERS

### A. Scenario

In open-access domestic broadband sharing schemes, where no trading benefits are associated with sharing, it is reasonable for *home users* to decide to share their connection based on the worst possible impact that they may experience. In this context, it would be useful from a scientific point-of-view to comprehend system dynamics in a worst-case-scenario situation, in which *guest users* request service at a time when *home users* make full use of their bandwidth.

To this end, we consider the following arbitrary but common scenario: A family of four people (i.e. *home users*) is sharing its broadband connection. Each *home user* is performing a different task online; *home user 1* uploads some photos to the cloud, *home user 2* holds a video call, *home user 3* plays an online game and *home user 4* browses the Internet. At the same time, an arbitrary number of *guest users* connect to the family access point and request web service.

### B. Setup

#### 1) Network topology and elements parameterization

Our simulations are based on the NS2 simulator [20]. The network topology and link characteristics depicted in Fig. 4 are used to represent the scenario described previously.

We consider two different access point profiles in our simulations: the first one (AP1) is based on a high-capacity fiber connectivity link, whereas the second one (AP8) is based on a conventional (i.e. medium-to-low capacity) ADSL link. Access point queue sizes are set to the respective 90$^{th}$ percentile value of each access point, as presented in Fig. 2.

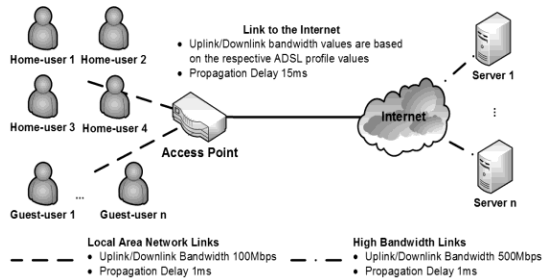

Fig. 4. Network topology used in the simulation campaign. Local area network and high-bandwidth links have the same properties over all simulations, whereas access point connectivity link to the Internet has different properties according to the respective access point profile used in each simulation.

In order to guarantee that the access point connectivity link to the Internet constitutes the sole bottleneck point in the network, each application is served by a different high-bandwidth network server.

#### 2) Packet scheduling configurations

Since we are interested in realistic and practical solutions, we consider only widely-used policies or policies that can be easily implemented based on existing QoS frameworks. Therefore, out of the vast amount of possible options, we consider the following packet scheduling configurations:

i) FIFO queueing, with a simple Drop Last (DL) dropping policy (DropTail). This configuration implies that the access point is simply left unlocked for *guest users* to use.

ii) FIFO queueing with different AQM dropping policies, including Random Early Detection (RED) and Controlled Delay (CoDel). In contrast to RED, which drops packets probabilistically based on predetermined queue size thresholds (*min_thresh*, *max_thresh*), CoDel checks periodically the average queueing delay and drops packets probabilistically when a predetermined queueing delay threshold is exceeded (*target delay*). The application of such dropping policies results in lower queueing delays and increased system fairness. In practice, we configure both RED and CoDel based on a *target delay* value of 5ms.

iii) Smoothed Round Robin fair queueing (SRR) [21]. SRR is an attractive fair queueing (FQ) packet scheduler, given its very low time complexity. With SRR, scheduling effort is evenly distributed across all flows requesting service.

iv) Strict non-preemptive priority queueing (PQ) implemented with two FIFO queues, one for each class of traffic. With this configuration, *guest users* are also allowed to request service from the access point in an unregulated manner, but unlike configurations (i), (ii) and (iii), home traffic has non-preemptive priority over guest traffic.

v) Regulated non-preemptive priority queueing as implemented by UPNQ. UPNQ parameters are selected based on the parameter values suggested in [13]. *UPNQ threshold* represents a queue percentage threshold for home-user packets, over which all newly arriving guest-user packets are forcefully dropped. As with configuration (iv), in UPNQ, home traffic has non-preemptive priority over guest traffic, but, additionally, guest traffic arrival rate is regulated, using the UPNQ percentage-based algorithm.

vi) Hybrid queueing as implemented by HPSS. We set *HPSS target delay impact* at 3ms. We consider this to be a reasonable impact value that *home users* can tolerate over their normal queueing delay. Both the *HPSS capacity threshold* and the amount of resources allocated to serve guest traffic are determined analytically as described in [15] based on the *target delay impact*. In particular, the corresponding *HPSS capacity threshold value* for a 3ms *target delay impact* is 2Mbps. Therefore, in access points with upload capacity below 2Mbps, such as AP8, HPSS behaves as a regulated non-preemptive priority queueing scheduler, using the HPSS temporal-based algorithm, while in access points with upload capacity above 2Mbps, such as AP1, HPSS behaves as a class-based WFQ scheduler. Within this context, authors of HPSS suggest for access points with *capacity thresholds* larger than 2Mbps, the allocation of 0.5% per Mbps of uplink bandwidth resources to serve exclusively guest-user traffic [15]; a suggestion that corresponds in practice to 3.25% of total uplink bandwidth resources for AP1 (i.e. 0.5 * 6.3 [Mbps] = 3.25).

vii) Class-based weighted fair queueing (CBQ), implemented with two FIFO queues, one for each class of traffic. CBQ is configured to allocate exclusively 95% and 5% of system resources for serving home and guest traffic, respectively. This particular configuration resembles commercial (e.g. FON's) broadband sharing services configurations, which typically allocate a relatively small but considerable percentage of broadband resources for serving exclusively guest traffic.

Finally, it should be noted that the evaluation of combination of policies, such as FQ-CoDel and WFQ-CoDel, is out of the scope of this study, as we are not interested in evaluating either the individual performance or fairness of flows within the home- and guest-user traffic classes.

### 3) Traffic characteristics

According to the scenario described earlier, home uplink traffic is composed by 2 elephants and 2 mice flows. Elephant flows are employed in this setting to represent the increased use of cloud syncing and video conferencing applications, which both produce large amounts of uplink traffic. Furthermore, although it would be more common for *home users* to engage in multiple network activities simultaneously, we argue that this particular home network setting represents a realistic traffic mix. Our argument is further supported by the employment of different types of network traffic, as well as by the simulation time, which is set to 10 minutes. More specifically, we use an FTP application to simulate the traffic generated from the first *home user*, a CBR application with 500Kbps bitrate to simulate the video call and an on/off traffic generator application, with an average bitrate of 12.8Kbps that creates a packet every 50ms, to simulate the updates sent to the game server [22]. Another traffic generator, which sends new web requests following *Lognormal* inter-arrival times [23], each with a size of 350bytes, is used to simulate the traffic generated from the fourth *home user*.

Unlike home-user traffic, our analysis results suggest that guest traffic is composed by a large number of mice flows. This type of traffic is generated by the same traffic generator application used in subsection V.D. In order to assess the scalability of the investigated packet scheduling configurations, we simulate a graduate volume increase of guest traffic (i.e. 1-3KBps, 6-8KBps, 13-15KBps, 44-46KBps), which is produced based on the four identified guest-user profiles.

### C. Results

Fig. 5 outlines the simulation results. The first row shows the results for AP1, while the second row shows the results for AP8. The following metrics are reported: i) guest-user average throughput, ii) guest traffic impact on home-user average throughput, iii) amount of guest-user data dropped and iv) guest traffic impact on home-user queueing delay. The impact values are calculated by the corresponding values attained by *home users* for the same packet scheduling configurations, with and without guest traffic, respectively.

The results of the lowest-bandwidth demand experiments (i.e. 1-3 KBps) show a minor impact of guest traffic. In particular, the impact on home-user average throughput remains below 1% across all configurations and for both access point profiles. The CBQ configuration in AP8 (see Fig. 5(d)) constitutes the sole exception to these observations, as the impact of guest traffic on home-user queueing delay reaches up to 4.5ms. This clearly indicates that a potential decision to exclusively allocate resources for guest traffic in systems with limited capacity might cause considerable impact on *home users* quality of experience, even when the actual guest traffic is extremely low.

The same set of results pertaining to the lowest-bandwidth demand experiments also show that there is a considerable deviation in the service ratios of guest traffic among the investigated configurations. In particular, for the high-capacity AP1 profile, all configurations, except from UPNQ, serve guest traffic adequately, and cause almost zero guest-user data drops. Due to the heavy network traffic generated by *home users*, UPNQ configuration follows a strict approach by allocating zero resources on guest traffic. For the medium-to-low capacity AP8 profile, HPSS and PRIO configurations resemble UPNQ, by following the same strict approach.

In contrast, with higher guest traffic loads, system behavior differs significantly depending on the employed packet scheduling configuration. The first conclusion that can be drawn is that, in the specific context of open-access domestic broadband sharing schemes, the DropTail configuration constitutes clearly an inappropriate choice. In particular, DropTail allows excessive amounts of guest traffic to be served – even when the network is congested - causing considerable impact on *home users* quality of experience. The impact grows further as the volume of guest traffic increases or/and the uplink bandwidth decreases, reaching up to 33%.

Interestingly, similar behavior is observed for three other configurations, which are based, respectively, on RED, CoDeL and SRR packet scheduling policies. These three policies penalize randomly flows belonging to both home and guest traffic in their effort to either enforce system fairness or preserve low queue latencies. Due to TCP's congestion-avoidance and control mechanisms, penalizing a home-user elephant flow forces the sender to significantly reduce its window size. In our scenario, such an action causes a severe degradation of the overall home-user network performance. Contrary to home traffic, penalizing a specific guest-user flow would not have the same devastating impact in the total volume of traffic originating from *guest users*, given that most guest-user packets belong to mice flows. Thus, independently of the penalty that AQM-based policies impose on guest-user flows, guest traffic will end up consuming a significant portion of system resources, as it is shown in Fig. 5. This behavior can be attributed to the incapability of these policies to differentiate flows based on traffic classes. This particular fact also implies that similar AQM-based schedulers, such as the Proportional Integral Controller Enhanced (PIE), would have presented analogous behavior.

Unlike Droptail, AQM- and FQ-based approaches, priority- and class-based packet scheduling approaches seem to balance more effectively the network requirements of *home* and *guest users*. They achieve that by serving a certain amount of guest traffic, without imposing large impact on *home users'* broadband performance. Among PRIO, UPNQ, HPSS and CBQ configurations, UPNQ configuration constitutes the strictest method, since it allocates the least amount of bandwidth for serving guest traffic. PRIO is the second strictest method. HPSS configuration presents the best overall performance across most cases, as it achieves a good balance between serving guest traffic and degrading home-user network performance. In particular, the imposed impact on home-user network performance is low, ranging from 0 up to 1.8%, regarding the average throughput, and less than 2.5ms, regarding queueing delay. CBQ configuration allocates the largest amount of network resources for serving guest traffic in comparison with all other priority- or class-based-queueing configurations. At the same time though, it imposes high queueing delay impact on *home users* network performance, which in our test cases, reaches up to 12ms.

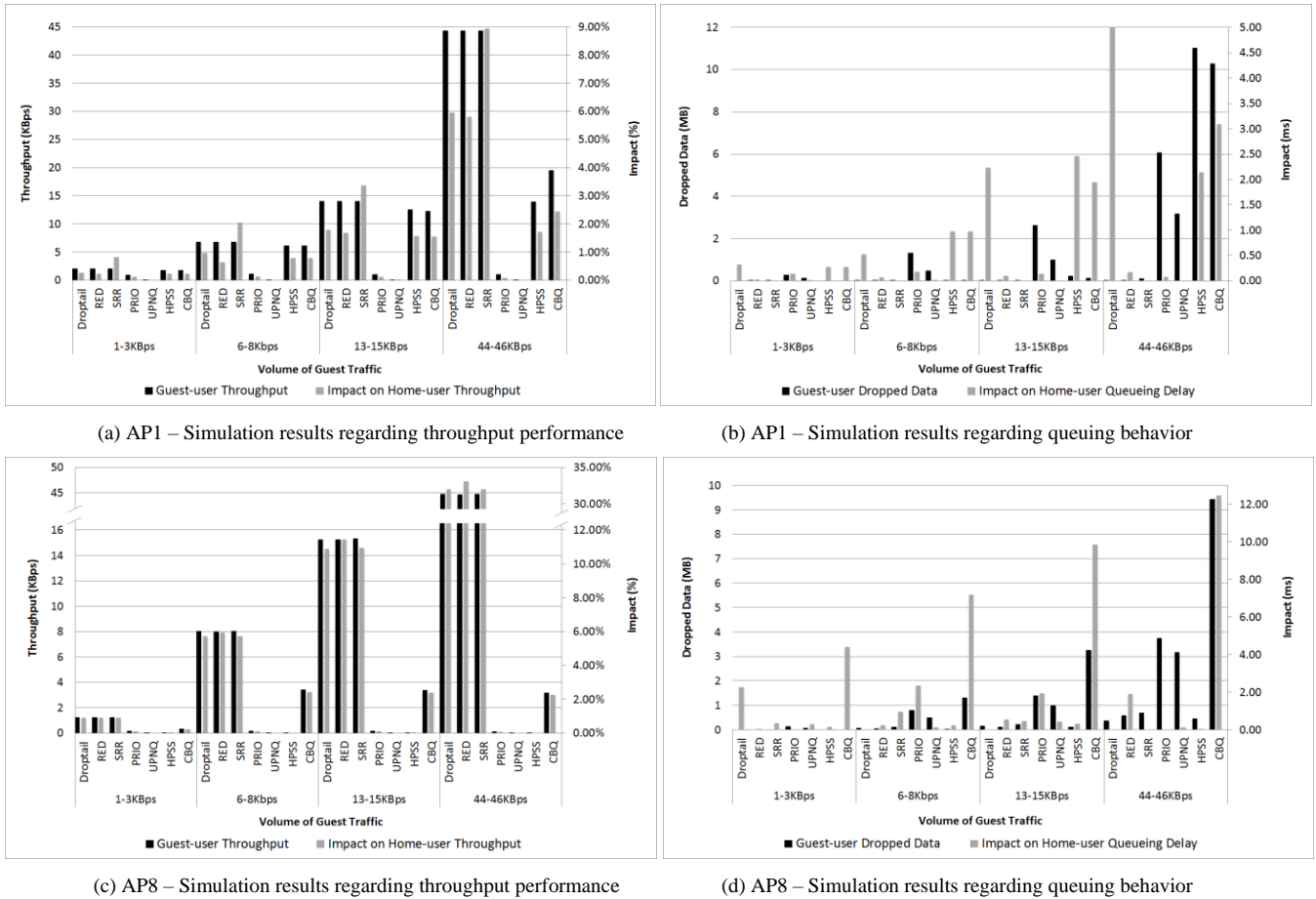

Fig. 5. Simulation results. Panels (a) and (c) present guest-user average throughput on the primary y-axis, with the respective impact values on home-user average throughput presented on the secondary y-axis. Panels (b) and (d) present the amount of dropped guest-user data on the primary y-axis, with the respective guest traffic impact on home-user queueing delay presented on the secondary y-axis. To increase the clarity of the figures, the results regarding CoDel have been omitted, due to their high similarity with the behavior presented by RED.

Balancing effectively the network requirements of *home* and *guest users* does not necessarily mean that *guest users* are always served adequately. As shown in Fig.5(c), configurations PRIO, UPNQ and HPSS allocate only a few bytes per second for serving guest traffic in their effort to confine the imposed impact on home-user network performance under tolerable levels. This particular observation demonstrates clearly the significant challenges associated with establishing an open-access domestic broadband sharing scheme, while employing a medium-to-low capacity broadband connection. Nevertheless, worst-case-scenario conditions investigated in this study are expected to occur rarely and/or for short timeframes in practice. Therefore, independently of the capacity of the broadband sharing network they have joined, *guest users* should be able to at least enjoy basic Internet services, such as email and chatting.

Finally, another aspect that should be carefully considered before selecting a packet scheduling configuration for home broadband sharing schemes is the overall dropping behavior with respect to guest traffic. Recent trends [24] suggest that smartphones and tablets will drive an even greater portion of future Internet traffic. We expect this percentage to be greater among the users of open-access Internet services, due to their opportunistic connectivity nature and the high availability of mobile devices. Packet scheduling configurations that cause large amounts of guest-user data to be dropped might affect the battery life of *guest users'* mobile devices negatively due to the excessive retransmission effort undertaken by clients to recover data losses. Our results reveal a trend in this topic, specifically regarding class-based-queueing approaches. This trend suggests that the more resources allocated by this type of queueing approaches for serving guest traffic, the largest the amount of data drops for *guest users*, given that guest-user resource demands are high. In particular, as more resources are allocated for serving guest traffic, guest-user queue will take more time to be filled up. In this time period, *guest users* will have further increased their transmission windows. Larger transmission windows result in more catastrophic data-loss events, once congestive collapse occurs.

## VII. CONCLUSIONS

In this paper, we assessed the functional capabilities of a wide set of packet scheduling configurations in effectively distributing open-access domestic broadband sharing network resources between *home* and *guest users*. Our results demonstrated clearly the inability of AQM- and FQ-based approaches to mitigate effectively the network requirements of *home* and *guest users* and instead showed that priority- and class-based-queueing approaches could serve better this purpose, by enabling *guest users* to access system resources in a less-than-best-effort manner.

Among these approaches, HPSS exhibited the most balanced behavior. In particular, in high-capacity broadband sharing networks, the amount of resources allocated by HPSS to serve guest traffic allowed *guest users* to enjoy a guaranteed level of service without hampering home-user network performance, whereas, in low-to-medium capacity broadband sharing networks, HPSS reduced significantly the amount of resources devoted to serve guest traffic by applying a strict priority queuing approach to protect *home users'* broadband quality of experience. These results confirm that both the selection of an appropriate packet scheduling policy for broadband sharing, as well as its respective configuration, depend largely on the throughput capacity of the broadband sharing connection; a view that was presented initially in [14].

Although hybrid packet scheduling approaches seem to be the most appropriate choice for broadband sharing schemes, our results also showed that class-based-queueing approaches with configurations that allocate more resources to serve guest traffic than that proposed by HPSS could also be considered as valid choices assuming that wider queueing delay tolerance levels are acceptable. It should be noted though, that such wider queueing delay levels are expected to be tolerated by *home users* only in cases where some form of reward policy for sharing is in place.

Furthermore, apart from the various performance aspects, our results also presented a trend with respect to the amount of resources allocated by class-based-queueing approaches for serving guest traffic and the data losses experienced by *guest users*. In particular, given that *guest users'* demand for bandwidth is high, our results showed that the more resources allocated to their service, the more severe data losses they experience. This trend is important as heavy losses may affect the battery life of *guest users'* mobile devices negatively and, therefore, lead to quality of experience degradation.

Overall, our work complements previous analytical and experimental studies on the topic [13] [15], and paves the way for the development of more sophisticated packet scheduling policies for open-access domestic broadband connection sharing schemes. In future work, we plan to expand the results presented in this paper with richer guest traffic datasets and investigate the potential use of delay tolerant networking concepts for handling guest traffic at gateways, under periods of heavy network usage from *home users*. In particular, our aim is to consider delay tolerant queue-management architectures for caching guest-user content at gateways and time-shifting its service in time periods, where home traffic is low.